\documentclass[11pt]{article}
\usepackage[pdftex]{graphicx}
\usepackage{caption}
\usepackage{subcaption}
\usepackage{dcolumn}% Align table columns on decimal point
\usepackage{bm}% bold math
\usepackage{color}
\usepackage{amsfonts}
\usepackage{amsmath}
\usepackage{stmaryrd}
\usepackage{epsfig}
\usepackage[english]{babel}
\usepackage[all]{xy}
\usepackage{ulem}
 \newtheorem{prop}{Proposition}

\usepackage[top=25.4mm, bottom=25.4mm, left=31.7mm, right=32.2mm]{geometry}
\begin{document}

\renewcommand{\baselinestretch}{1.25}

\title{B\"{a}cklund-Darboux Transformations and Discretizations of $N=2\; a=-2$ Supersymmetric KdV Equation}

\author{Hui Mao and Q. P. Liu\footnote{Corresponding author. Email: qpl@cumtb.edu.cn}\\
Department of Mathematics, \\
China University of Mining and Technology,\\
Beijing 100083, People's Republic of China\\[5pt]
}

\date{}
\maketitle

\begin{abstract}
The $N=2 \;a=-2$ supersymmetric KdV equation is  studied. A Darboux transformation and the corresponding B\"acklund transformation are constructed for this equation.
 Also, a nonlinear superposition formula is worked out
for the associated  B\"acklund transformation. The B\"{a}cklund transformation and the related nonlinear superposition formula are used to construct integrable super semi-discrete and full discrete systems. The  continuum limits of these discrete systems  are also considered.
\end{abstract}

{\bf Key words:} {Supersymmetric integrable system, B\"acklund transformation, Darboux transformation, nonlinear superposition formula, discrete integrable system}
\maketitle %\maketitle must follow title, authors, abstract and

\section{Introduction}
The Korteweg-de Vries (KdV) equation probably is the most important equation in the theory of modern integrable systems or soliton theory. Indeed, it is well known that the theory originated from the celebrated work of Kruskal and his collaborators on the KdV equation. Due to its remarkable properties and wide applications, the KdV equation has been extended in various ways and one of them is the supersymmetric extension.

In the theory of supersymmetric integrable systems, there are two classes of extensions, namely the nonextended ($N=1$) and the extended ($N>1$) extensions. It is interesting to note that in the extended situation new bosonic systems may appear as a bonus. For the KdV equation, its nonextended supersymmetric version was introduced by Manin and Radul \cite{MR} and its extended ($N=2$) version by Laberge and Mathieu \cite{LCM}. The $N=1$ supersymmetric KdV equation has been studied extensively and various properties have been established for it (see \cite{MP1}-\cite{lou2} and the references there).

%itself has been developed
%extensively since eighties of last century.  As the result of this development, many supersymmetric integrable equations have been
%studied and a number of interesting properties has been established.
%Among them, the most celebrated supersymmetric system is the
%supersymmetric Korteweg-de Vries (KdV) equation \cite{MR,MP1}. It
%has been shown that, as its bosonic analogue, the supersymmetric KdV
%(SKdV) equation is a bi-Hamiltonian system \cite{OP1}, has Darboux
%and B\"{a}cklund transformations \cite{LM,LX}, can be casted into
%bilinear from \cite{MY,CAS,CRG}, etc.
%In the literature, there exist more than one supersymmetric
%extensions for the KdV equation. The most interesting ones are the
%N=2 supersymmetric KdV equations.  The system was originally
%introduced by Laberge and Mathieu \cite{LCM,LPM}.
The $N=2$ supersymmetric KdV system reads as \cite{LCM}
\begin{equation}\label{SKdVa}
\phi_t=-\phi_{xxx}+3(\phi{\cal D}_1{\cal
D}_2\phi)_x+\frac{1}{2} (a-1)({\cal D}_1{\cal
D}_2\phi^2)_x+3a\phi^2\phi_x,
\end{equation}
where $\phi=\phi(x,t,\theta_1,\theta_2)$ is a superboson function depending on temporal variable $t$, spatial variable $x$ and its fermionic counterparts $\theta_i (i=1,2)$. ${\cal D}_1$ and ${\cal D}_2$ are the super derivatives defined by ${\cal D}_1=\partial_{\theta_1}+\theta_1\partial_x, {\cal D}_2=\partial_{\theta_2}+\theta_2\partial_x$ and $a$ is a free parameter. In the sequel, we will refer to (\ref{SKdVa}) as the SKdV$_a$ equation. It is known that this one-parameter family of equations (\ref{SKdVa}) is integrable only for certain values of the parameter $a$. For both $a=-2$ and $a=4$ cases, Laberge and Mathieu in \cite{LCM} demonstrated that the systems possess Lax representations and Hamiltonian structures and
infinite conservation laws. For the SKdV$_1$ equation, its Lax representation was constructed by Popowicz in \cite{Pop}. While the bi-Hamiltonian structure of SKdV$_4$ equation was found by  Kupershmidt (see ref. \cite{LPM}), a systematic construction of bi-Hamiltonian structures both for  SKdV$_4$ and SKdV$_{2}$ equations is due to Oevel and Popowicz \cite{OP1} within the framework of $r$-matrix theory (see also \cite{figue}). Later, a recursion operator and a bi-Hamiltonian structure for  SKdV$_1$ equation were obtained by Sorin and Kersten \cite{sorin}. Bourque and Mathieu, through Painlev\'e analysis, showed that above three cases pass the test and confirmed their integrability \cite{BM}.
%We also remark that $N=2$ SKdV systems can be represented in terms of$N=1$ Lax operators \cite{IK,Liu2}.

The $N=2$ SKdV systems were studied from the viewpoint of Hirota bilinear method and both SKdV$_4$ and SKdV$_1$ systems were successfully brought to Hirota bilinear forms \cite{zhang}. Then, for   SKdV$_4$ equation, a class of solutions was calculated and for  SKdV$_1$ equation, a bilinear B\"acklund transformation was obtained. To the best of our knowledge, a proper Hirota bilinear representation of the SKdV$_{-2}$ equation is still missing, although some attempts have been made by Delisle and Hussin \cite{Delisle}\cite{hussin} and certain interesting solutions have been obtained. Also, any form of B\"acklund transformation is not known  to this equation.

The main purpose of this paper is to present a Darboux transformation and the related B\"{a}cklund  transformation for the SKdV$_{-2}$ equation.
 The theory of Darboux and B\"{a}cklund transformations has been an integrated part of the soliton theory \cite{matveev,rogers, Gu,cies,dl} and it is important for a given nonlinear system to find its B\"acklund transformations. With a B\"acklund transformation in hand, one may either construct various solutions for the associated nonlinear system or produce new integrable systems of both continuous and discrete types \cite{levi1980,levi1981,frank}. It is noted that B\"acklund transformations for the supersymmetric integrable systems emerged already as early as the later seventies of last century \cite{kulish}, their applications to integrable discretizations of super integrable systems were developed only recently \cite{sasha,xll,xl,xue,carstea2015}. Very recently, the relationship between these results and extensions of Yang-Baxter map was explored \cite{sasha1, sasha2}.

The paper is organized as follows. In  Section 2 we recall the Lax representation  for the SKdV$_{-2}$ equation  and derive its  Darboux and B\"{a}cklund transformation through gauge transformation. In  Section 3 we obtain nonlinear superposition formula from the associated  B\"acklund transformation. Then in Section 4 we use the obtained transformations to construct discrete integrable super systems. Both differential-difference equations and difference-difference equations are obtained. In Section 5, by performing various continuum limits, we show that our discrete systems are indeed the proper discretizations of the SKdV$_{-2}$ equation.

\section{B\"{a}cklund-Darboux Transformations}
Our SKdV$_{-2}$ equation may be read from  (\ref{SKdVa}), namely
\begin{equation}
\phi_t=\left(-\phi_{xx}+3(\phi{\cal D}_1{\cal
D}_2\phi)-\frac{3}{2}({\cal D}_1{\cal D}_2\phi^2)-2\phi^3\right)_x,
\label{SKdV-2}
\end{equation}
to rewrite it in $N=1$ formalism, we assume  \[\phi=v+\theta_{2}\alpha,\]
where $v=v(t,x,\theta_1)$ is a bosonic (even) function while
$\alpha=\alpha(t,x,\theta_1)$ is a fermionic (odd) one. Then the
SKdV$_{-2}$ equation (\ref{SKdV-2}) in components takes the following form
\begin{subequations}\label{eq1}
 \begin{align}
  &v_{t}=(-v_{xx}-3(\mathcal{D}v)\alpha-2v^{3})_{x},\\
  &\alpha_{t}=(-\alpha_{xx}+3\alpha(\mathcal{D}\alpha)+3(\mathcal{D}v)v_{x}-6v^{2}\alpha)_{x},
 \end{align}
\end{subequations}
where for simplicity we replace ${\mathcal D}_1$ by ${\mathcal{D}}$.

Through the following invertible change of variables
\[
\alpha=\beta-\frac{1}{2}\mathcal{D}u, \quad v=-\frac{i}{2}u,
\]
where $u=u(t,x,\theta_1)$ is a bosonic (even) function and
$\beta=\beta(t,x,\theta_1)$ is a fermionic (odd) one and $i=\sqrt{-1}$, the system \eqref{eq1} is brought to
\begin{subequations}\label{eq3}
 \begin{align}
  &u_{t}=\left(-u_{xx}-3(\mathcal{D}u)\beta+\frac{1}{2}u^{3}\right)_{x},\\
  &\beta_{t}=\left(-\beta_{xx}+3\beta(\mathcal{D}\beta)-3u_{x}\beta+\frac{3}{2}u^{2}\beta\right)_{x}.
 \end{align}
\end{subequations}
Above system is nothing but the system constructed by Inami and Kanno based on the affine Lie super algebra $A(1,1)^{(1)}$ \cite{IK}.
Therefore, rather than \eqref{eq1} we shall refer the system \eqref{eq3} as the SKdV$_{-2}$ and work with it in the subsequent discussion.

According to Inami and Kanno, equation (\ref{eq3}) possesses a Lax representation \cite{IK}
\[
L_t=[P, L],
\]
where
\begin{equation}\label{eqlax}
   L = \partial_{x}^2-u\partial_{x}-\beta\mathcal{D},
\end{equation}
and
\[
 \quad P = -4\partial_{x}^3+6u\partial_{x}^2+6\beta\mathcal{D}^3+(3u_{x}-\frac{3}{2}u^2)\partial_{x}+(3\beta_{x}-3u\beta)\mathcal{D}.
\]
%= -4(L^{\frac{3}{2}})_{\ge1}
Thus, the corresponding linear spectral problem is
\begin{equation}\label{eigen}
L\phi=\lambda\phi.
\end{equation}
One well adopted way to find a Darboux transformation is the gauge transformation approach. To do so, we first reformulate the linear spectral problem \eqref{eigen} into the matrix form. Introducing $\Phi=(\phi,\phi_{x},\mathcal{D}\phi,\mathcal{D}\phi_{x})^T $,  we may rewrite (\ref{eigen}) as
\begin{eqnarray}\label{eq:4}
\mathcal{D}\Phi=M\Phi,\quad M=\left(
    \begin{array}{cccc}
      0 & 0 & 1 & 0 \\
      0 & 0 & 0 & 1 \\
      0 & 1 & 0 &0 \\
      \lambda & u &\beta & 0 \\
    \end{array}
  \right).
\end{eqnarray}

The matrix $M$  is a super matrix, by which we mean matrix with entries involving both bosonic and fermionic variables.
Following \cite{xue},  we introduce an involution on the algebra of super matrices as follows: given
any matrix $A=(a_{ij})_{i,j\in \mathbb{Z}}$, we define  $A^\dag=(a_{ij}^\dag)_{i,j\in \mathbb{Z}}$
and $a_{ij}^\dag=(-1)^{p(a_{ij})}a_{ij}$  with $p(a_{ij})$ denoting the parity of $a_{ij}$.

To construct a Darboux transformation for \eqref{eq:4}, we seek for a gauge matrix $T$ such that
\begin{equation}\label{eq:5}
  \Phi_{[1]}=T\Phi
\end{equation}
 solves
\begin{equation}\label{eq:6}
  \mathcal{D}\Phi_{[1]}=M_{[1]}\Phi_{[1]},
\end{equation}
where $M_{[1]}$ is the matrix $M$ but with $u, \beta$ replaced by the new field variables $u_{[1]}, \beta_{[1]}$.
A qualified Darboux transformation requires  the gauge matrix $T$  to satisfy
\begin{equation}\
  \mathcal{D}T+T^{\dag}M-M_{[1]}T=0.
\end{equation}
Our main aim now is to find a solution to above equation, thus we make the following ansatz, namely
 \[
 T=\lambda F+G, \; F=(f_{ij})_{4\times 4}, \;  G=(g_{ij})_{4\times 4}.
 \]
Through tedious calculations, we find that the matrices $F$ and $G$ may be taken as
\begin{equation}\label{feq}
F=\begin{pmatrix}0&0&0&0\\ a&0&0&0\\ 0&0&0&0\\ \mathcal{D}a-a\eta&0&a&0\end{pmatrix},
\end{equation}
and
\begin{equation}\label{geq}
G=\left(\begin{array}{cccc}
  \lambda_{1} & a & a\eta & 0\\
  0&  a^2-(\mathcal{D}a)\eta & (\lambda_{1}+a^2+a\mathcal{D}\eta)\eta & a\eta\\
  0& \mathcal{D}a-a\eta & \lambda_{1}+a\mathcal{D}\eta+(\mathcal{D}a)\eta  & a\\
  0&  (2a+\mathcal{D}\eta)(\mathcal{D}a-a\eta) & \lambda_{1}\mathcal{D}\eta+a(\mathcal{D}a)\eta+(a+\mathcal{D}\eta)\mathcal{D}(a\eta) &a(a+\mathcal{D}\eta)
\end{array}\right).
\end{equation}
It is noted that all the entries of the Darboux matrix are given in terms of two basic quantities $a$ and $\eta$. Now direct calculation shows that  $T$ is  a Darboux matrix provided  that  $a$ and $\eta$  satisfy the following equations
\begin{subequations}\label{bt}
 \begin{align}
  &a_{x}=a^2-au-\lambda_{1}-(\mathcal{D}a)\eta,\\
  &\eta_{x}=\left(u+\frac{2\lambda_{1}}{a}+\mathcal{D}\eta\right)\eta-\beta,
 \end{align}
\end{subequations}
and the transformations between field variables are neatly given by
\begin{equation}\label{dtpotential}
  u_{[1]}=u+\frac{2a_x}{a}, \quad \beta_{[1]}=\beta+2\eta_x.
\end{equation}

We may obtain a B\"acklund transformation (spatial part) for the SKdV$_{-2}$ equation \eqref{eq3} from the equations \eqref{bt} and \eqref{dtpotential} via elimination of $a$ and $\eta$. A convenient way to achieve the goal is to introduce new variables via
\[
u=2(\ln v)_x, \; u_{[1]}=2(\ln v_{[1]})_x,  \;  \beta=2\gamma_x, \; \beta_{[1]}=2\gamma_{[1],x},
\]
and then system \eqref{eq3} is brought to
 \begin{subequations}\label{eqpotential}
 \begin{align}
  &v_{t}=\left(-(\frac{v_{x}}{v})_{xx}-6\mathcal{D}(\frac{v_{x}}{v})\gamma_{x}+2(\frac{v_{x}}{v})^3\right)v,\\
  &\gamma_{t}=-\gamma_{xxx}+6\gamma_{x}\mathcal{D}(\gamma_{x})-6\gamma_{x}(\frac{v_{x}}{v})_{x}
   +6\gamma_{x}(\frac{v_{x}}{v})^2,
 \end{align}
\end{subequations}
which basically is the potential form of \eqref{eq3}. Now from \eqref{dtpotential} we have
\[
a=k_{1}\frac{v_{[1]}}{v}, \quad  \eta=\gamma_{[1]}-\gamma,
\]
where $k_{1}$ is a constant of integration. Substituting them into \eqref{bt} and letting  $\lambda_1=k_{1}^2$, we reach the following B\"acklund transformation
\begin{subequations}\label{btc}
 \begin{align}
   (v_{[1]}v)_{x}&=k_{1}(v_{[1]}^2-v^2)+(\gamma_{[1]}-\gamma)(v\mathcal{D}v_{[1]}-v_{[1]}\mathcal{D}v),\\
   (\gamma_{[1]}+\gamma)_{x}&=\left(\frac{2v_{x}}{v}+2k_{1}\frac{v}{v_{[1]}}+\mathcal{D}\gamma_{[1]}-\mathcal{D}\gamma\right)(\gamma_{[1]}-\gamma).
 \end{align}
\end{subequations}

The Darboux matrix $T$ is determined by $\lambda_1$, $a$ and $\eta$ and it is observed that these quantities may be related to the solutions of the linear spectral problem \eqref{eq:4} in such way that the Darboux matrix $T$ may be made more explicit.  This can be done by studying the kernel of the Darboux matrix $T$. To this end, we take  two particular solutions $\Phi_0=(\phi_0,\phi_{0x},\mathcal{D}\phi_0,\mathcal{D}\phi_{0x})^T $ and $\Phi_1=(\phi_1,\phi_{1x},\mathcal{D}\phi_1,\mathcal{D}\phi_{1x})^T $ of \eqref{eq:4} at $\lambda=\lambda_1$ and then  we  find
\begin{subequations}\label{aeta}
 \begin{align}
  &a=-\frac{\lambda_1}{\phi_{0x}}\left(\phi_{0}-\frac{\phi_{1}\mathcal{D}\phi_{0}}{\mathcal{D}\phi_{1}}
     +\frac{\phi_{0}\phi_{1x}\mathcal{D}\phi_{0}}{\phi_{0x}\mathcal{D}\phi_{1}}\right),\\
  &\eta=-\frac{1}{{\mathcal D}\phi_1}\left(\phi_{1x}-\frac{\phi_{1}\phi_{0x}}{\phi_{0}}
     +\frac{\phi_{1}\phi_{1x}\mathcal{D}\phi_{0}}{\phi_{0}(\mathcal{D}\phi_{1})}\right),
 \end{align}
\end{subequations}
where $\phi_{0}$ is a bosonic function while $\phi_{1}$ is a fermionic one.

Summarizing above discussions, we have
\begin{prop}
Let $\phi_{0}$ and $\phi_{1}$ be two solutions of the linear spectral problem \eqref{eigen} at $\lambda=\lambda_1$, the first one being bosonic and the second one fermionic. Let the matrices $F$ and $G$  be given by \eqref{feq} and \eqref{geq} with $a$ and $\eta$ defined by \eqref{aeta}. Then $T=\lambda F+G$ is a Darboux matrix for the linear spectral problem \eqref{eq:4}. The field variables transform according to \eqref{aeta}.
\end{prop}
Concluding this section, we mention that the corresponding Darboux transformation for the scalar spectral problem \eqref{eigen} may be obtained from above result, namely
\[
\phi[1]=\lambda_1\phi+a(\partial +\eta {\mathcal D})\phi.
\]
\section{Nonlinear Superposition Formula}
B\"acklund transformation is interesting and important because it may provide a way to construct solutions to a given nonlinear system. However,  B\"acklund transformation  itself is a system of differential equations, therefore it may not be easy  to find explicit solutions. One way to get over this difficulty is to derive the corresponding nonlinear superposition formula.
 %available, For a given B\"{a}cklund transformation, it is desirable to find the corresponding nonlinear superposition formula since on the one hand such formula often provides a convenient way to construct solutions to the nonlinear system, it may also supply a (semi-) discrete system on the other hand.

In the present case, to find the nonlinear superposition formula of B\"{a}cklund transformation we obtained in last section, we suppose that
($v,\gamma$) is an arbitrary solution of the potential SKdV$_{-2}$ \eqref{eqpotential}, then with the help of two B\"acklund parameters $\lambda_{j}(j=1,2)$, we may get the new solution $v_{[j]},\gamma_{[j]}$ and $\Phi_{[j]}=T|_{\lambda=\lambda_{j}}\Phi$. That is, we consider a pair of Darboux transformations
\begin{align}
&\Phi_{[1]}=T_{[1]}\Phi, \;\;\; T_{[1]}\equiv T|_{\lambda_1=k_{1}^2, a=a_1,\eta=\eta_1}\label{phi1},&\\
&\Phi_{[2]}=T_{[2]}\Phi, \;\;\; T_{[2]}\equiv T|_{\lambda_1=k_{2}^2, a=a_2,\eta=\eta_2},&\label{phi2}
\end{align}
then with the help of the Bianchi's permutability theorem,  depicted by the diagram below
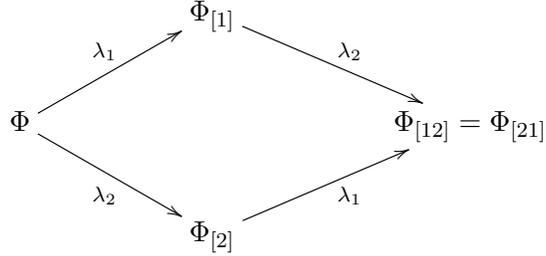
\begin{displaymath}
     \xymatrix{
                                & &\Phi_{[1]}\ar[drr]^{\lambda_2}&&\\
         \Phi\ar[urr]^{\lambda_1} \ar[drr]_{\lambda_2}  &  &&&\Phi_{[12]}= \Phi_{[21]}
         \\
                             && \Phi_{[2]}\ar[urr]_{\lambda_1}&& }
\end{displaymath}
we have
\begin{equation}\label{eq5}
  T_{[12]} T_{[1]} =T_{[21]} T_{[2]},
\end{equation}
where
\[
 T_{[12]}\equiv T|_{\lambda_1=k_{2}^2, a=a_{12},\eta=\eta_{12}}, \quad
  T_{[21]}\equiv T|_{\lambda_1=k_{1}^2, a=a_{21},\eta=\eta_{21}},
\]
\[a_{12}=k_{2}\frac{v_{[12]}}{v[1]},\quad a_{21}=k_{1}\frac{v_{[21]}}{v[2]},
\quad \eta_{12}=\gamma_{[12]}-\gamma_{[1]}, \quad \eta_{21}=\gamma_{[21]}-\gamma_{[2]}.\]
From $v_{[12]}=v_{[21]},\gamma_{[12]}=\gamma_{[21]}$ and \eqref{eq5}, after some tedious calculations we obtain for SKdV$_{-2}$ equation the following nonlinear superposition formula
\begin{subequations}\label{nsf}
 \begin{align}
  \frac{v_{[12]}}{v}&=\frac{k_{1}v_{[2]}-k_{2}v_{[1]}}{k_{1}v_{[1]}-k_{2}v_{[2]}}
   +\frac{(k_{1}^2-k_{2}^2)(\gamma_{[1]}-\gamma_{[2]})
     (v_{[2]}\mathcal{D}v_{[1]} -v_{[1]}\mathcal{D}v_{[2]})}
     {(k_{1}v_{[1]}-k_{2}v_{[2]})^2\left(\mathcal{D}\gamma_{[1]}-\mathcal{D}\gamma_{[2]}
     +\frac{k_1v}{v_{[1]}}-\frac{k_2v}{v_{[2]}}\right)},\\
  \gamma_{[12]}-\gamma&=\frac{(k_{1}^2-k_{2}^2)(\gamma_{[2]}-\gamma_{[1]})v_{[1]}v_{[2]}}
     {v(k_{1}v_{[2]}-k_{2}v_{[1]})\left(\mathcal{D}\gamma_{[1]}-\mathcal{D}\gamma_{[2]}
     +\frac{k_1v}{v_{[1]}}-\frac{k_2v}{v_{[2]}}\right)}.
 \end{align}
\end{subequations}
As in most cases, the nonlinear superposition formula is of differential-algebraic type.

\section{Discrete equations}

Both B\"acklund transformation \eqref{btc} and the nonlinear superposition formula \eqref{nsf} may be interpreted as (semi-) discrete systems.
To this end,  we rewrite the super fields in terms of its components. Let us assume
\[
v=p+\theta_1 \rho, \;\; \gamma=\sigma+\theta_1 q
\]
and put them into (\ref{eqpotential}), SKdV$_{-2}$ equation takes the following component form
\begin{subequations}\label{skdvcomponent}
 \begin{align}
  pp_{t}=&-pp_{xxx}+3p_{x}p_{xx}+6(p_{x}\rho-p\rho_{x})\sigma_{x},\\
  (p\rho)_{t}=&-(p\rho)_{xxx}+6(p_{x}\rho_{xx}+\rho_{x}p_{xx})
  +6(p_{x}^2-pp_{xx}+2\rho_{x}\rho)\sigma_{x}-6q_{x}(p_{x}\rho-p\rho_{x}),  \\
  \sigma_{t}=&-\sigma_{xxx}+6(q_{x}-\frac{p_{xx}}{p}+\frac{2p_{x}^2}{p^2})\sigma_{x},\\
  q_{t}=&-q_{xxx}+6q_{x}(q_{x}-\frac{p_{xx}}{p}+\frac{2p_{x}^2}{p^2})
  -6\sigma_{x}(\sigma_{xx}-\frac{\rho_{xx}}{p}+\frac{\rho p_{xx}}{p^2}
  +4\frac{p_{x}\rho_{x}}{p^2}-4\frac{p_{x}^2\rho}{p^3}).
 \end{align}
\end{subequations}
Define
\begin{eqnarray}\nonumber
  &\ p\equiv p_{n}(x),\quad \;p_{[1]}\equiv p_{n+1}(x),\;\quad \rho\equiv \rho_{n}(x),\quad\; \rho_{[1]}\equiv \rho_{n+1}(x),\\\nonumber
  &\ q\equiv q_{n}(x),\quad\; q_{[1]}\equiv q_{n+1}(x),\;\quad \sigma\equiv \sigma_{n}(x),\quad \;\sigma_{[1]}\equiv \sigma_{n+1}(x).
\end{eqnarray}
Then, the  B\"{a}cklund transformation \eqref{btc} is split into the system
\begin{subequations}\label{btdiscrete}
 \begin{align}
(p_{n+1}p_{n})_x=&\;
  k_{1}(p_{n+1}^2-p_{n}^2)+(\sigma_{n+1}-\sigma_{n})(p_{n}\rho_{n+1}-p_{n+1}\rho_{n}),  \\
 (p_{n+1}\rho_{n}+\rho_{n+1}p_{n})_x=&\;2k_{1}(p_{n+1}\rho_{n+1}-p_{n}\rho_{n})+(q_{n+1}-q_{n})(p_{n}
 \rho_{n+1}-p_{n+1}\rho_{n})  \nonumber\\
 &-(\sigma_{n+1}-\sigma_{n})(p_{n} p_{n+1,x}-p_{n+1}p_{n,x}+2\rho_{n}\rho_{n+1}),  \\
 (\sigma_{n+1}+\sigma_{n})_x=&\;\left(\frac{2p_{n,x}}{p_{n}}+\frac{2k_{1}p_{n}}{p_{n+1}}+q_{n+1}-q_{n}\right)(\sigma_{n+1}-\sigma_{n}),\\
 (q_{n+1}+q_{n})_x=&\;\left((\frac{2\rho_{n}}{p_{n}})_x
  +\frac{2k_{1}}{p_{n+1}}(\rho_{n}-\frac{p_{n}\rho_{n+1}}{p_{n+1}})+(\sigma_{n+1}-\sigma_{n})_x\right)(\sigma_{n+1}-\sigma_{n})\nonumber\\
 &+\left(\frac{2p_{n,x}}{p_{n}}+\frac{2k_1p_{n}}{p_{n+1}}+q_{n+1}-q_{n}\right)(q_{n+1}-q_{n}),
 \end{align}
\end{subequations}
which  is a semi-discrete or differential-difference system.

To find a difference-difference system,    we  define for any field variable $u$
\[
u\equiv u_{n,m},\quad \;u_{[1]}\equiv u_{n+1,m},\;\quad u_{[2]}\equiv u_{n,m+1},\;\quad u_{[12]}\equiv u_{n+1,m+1},
\]
%\begin{align*}
%p &\equiv  p_{n,m},& p_{[1]}&\equiv p_{n+1,m}, &\quad p_{[2]}&\equiv p_{n,m+1},& p_{[12]}&\equiv p_{n+1,m+1},\\
%\rho & \equiv  \rho_{n,m},&\rho_{[1]}&\equiv \rho_{n+1,m},& \rho_{[2]}&\equiv \rho_{n,m+1},&\rho_{[12]}&\equiv \rho_{n+1,m+1},\\
%q &\equiv  q_{n,m}, & q_{[1]}&\equiv q_{n+1,m},& q_{[2]}&\equiv q_{n,m+1},& q_{[12]}&\equiv q_{n+1,m+1},\\
%\sigma &\equiv  \sigma_{n,m},&\sigma_{[1]}&\equiv \sigma_{n+1,m},&\sigma_{[2]}&\equiv \sigma_{n,m+1},&\sigma_{[12]}&\equiv \sigma_{n+1,m+1}.
%\end{align*}
then the nonlinear superposition formula  \eqref{nsf} yields
\begin{subequations}\label{nsfdiscrete}
{\setlength\arraycolsep{1.5pt}\begin{eqnarray}
 \frac{p_{n+1,m+1}}{p_{n,m}}+\frac{a(p;1,2)}{a(p;2,1)}&=&\frac{(k_1^2-k_2^2)b(\sigma,p,p)b(\rho/p,p,p)}{a^2(p;2,1)
 (p_{n,m}a(p;1,2)-b(q,p,p))}, \\
 \frac{p_{n,m}\rho_{n+1,m+1}-p_{n+1,m+1}\rho_{n,m}}{(k_1^2-k_2^2)p_{n,m}^2}&=&\frac{b(\rho/p,p,p)a(p;1,2)-b(\rho,p^2,p^2)\left(\ln\frac{p_{n,m+1}}{p_{n+1,m}}\right)_x
 }{a^2(p;2,1)(p_{n,m}a(p;1,2)-b(q,p,p))}\nonumber\\
& &+\frac{b(\sigma,p,p)b(\rho/p,p,p)(b(\sigma_x,p,p)-\rho_{n,m}a(p;1,2))}{a^2(p;2,1)(p_{n,m}a(p;1,2)-b(q,p,p))^2} \nonumber \\
& &+\frac{a(p;1,2)b(\sigma,\rho p,\rho p)}{a^2(p;2,1)(p_{n,m}a(p;1,2)-b(q,p,p))^2},\\
 \frac{p_{n,m}\left(\sigma_{n+1,m+1}-\sigma_{n,m}\right)}{k_1^2-k_2^2}&=&\frac{b(\sigma,p^2,p^2)}{a(p;1,2)(p_{n,m}a(p;1,2)-b(q,p,p))^2}, \\
\frac{(q_{n+1,m+1}-q_{n,m})a(p;1,2)}{k_1^2-k_2^2}&=&\frac{b(q,p^2,p^2)+(\frac{\rho_{n,m+1}}{p_{n,m+1}}+\frac{\rho_{n+1,m}}{p_{n+1,m}})b(\sigma,p^2,p^2)}{p_{n,m}(p_{n,m}a(p;1,2)-b(q,p,p))}\nonumber\\
&&+\frac{\rho_{n,m}a(p;1,2)b(\sigma,p^2,p^2)-b(\sigma,p,p)b(\sigma_x,p^2,p^2)}{p_{n,m}(p_{n,m}a(p;1,2)-b(q,p,p))^2}\nonumber\\
&&+\frac{(\rho_{n,m}a(p;1,2)+p_{n,m}a(\rho;1,2))p_{n+1,m}p_{n,m+1}}{p^2_{n,m}a(p;1,2)(p_{n,m}a(p;1,2)-b(q,p,p))}\nonumber\\
&&+\frac{b(\sigma,p^3,p^3)a(\rho/p;2,1)}{(p_{n,m}a(p;1,2)-b(q,p,p))^2},
\end{eqnarray}}
\end{subequations}
where
%\newpage
\begin{align*}
a(p;1,2)&\equiv k_1 p_{n,m+1}-k_2p_{n+1,m}£¬ & b(p,q,r)&\equiv (p_{n,m+1}-p_{n+1,m})q_{n+1,m}r_{n,m+1},
\end{align*}

In the present form, above system (24) is a semi-discrete one. It is interesting to note that a fully discrete system may be obtained if the following relation, which comes from B\"{a}cklund transformation, is taken into account
\begin{subequations}\label{extra}
\begin{align}
\frac{p_{n,m+1,x}}{p_{n,m+1}}-\frac{p_{n+1,m,x}}{p_{n+1,m}}=&\;\frac{k_2p_{n,m+1}}{p_{n,m}}
-\frac{k_1p_{n+1,m}}{p_{n,m}}+p_{n,m}\left(\frac{k_1}{p_{n+1,m}}-\frac{k_2}{p_{n,m+1}}\right)\nonumber\\
&+(\sigma_{n,m+1}-\sigma_{n+1,m})\left(\frac{\rho_{n,m+1}}{p_{n,m+1}}-\frac{\rho_{n,m}}{{p_{n,m}}}\right)  \nonumber \\
&+(\sigma_{n+1,m}-\sigma_{n,m})\left(\frac{\rho_{n,m+1}}{p_{n,m+1}}-\frac{\rho_{n+1,m}}{p_{n+1,m}}\right),\\
(\sigma_{n+1,m}-\sigma_{n,m+1})_x=&\;\left(\frac{2p_{n,m,x}}{p_{n,m}}+\frac{2k_{1}p_{n,m}}{p_{n+1,m}}
+q_{n+1,m}-q_{n,m}\right)(\sigma_{n+1,m}-\sigma_{n,m+1}) \nonumber \\
  &+\left(\frac{2k_{1}p_{n,m}}{p_{n+1,m}}-\frac{2k_{2}p_{n,m}}{p_{n,m+1}}+q_{n+1,m}-q_{n,m+1}\right)(\sigma_{n,m+1}-\sigma_{n,m}).
\end{align}\end{subequations}

%Put them into (\ref{nsfdiscrete}),we can eliminate the derivative terms and obtain a difference-difference system.

\section{Continuum limits}
In the last section, with the help of the B\"acklund transformation and the related nonlinear superposition formulae, we obtained a semi-discrete system and a fully discrete system. It is interesting to identify these discrete systems and analyze  their continuum limits \cite{frank}. We will show that the obtained systems (\ref{btdiscrete}) and  (\ref{nsfdiscrete}) are nothing but the discrete versions of the potential SKdV$_{-2}$ (\ref{skdvcomponent}) in component form.

%\subsection{The continuous limit of (\ref{btdiscrete})}
Let us first consider (\ref{btdiscrete}). Introducing  the new continuous variable $\tau$ by
\begin{align*}
p_{n}(x)\equiv p(x,\tau),\rho_{n}(x)\equiv \rho(x,\tau),
q_{n}(x)\equiv q(x,\tau),\sigma_{n}(x)\equiv \sigma(x,\tau),\tau=\frac{n}{k_1},
\end{align*}
and expanding
\begin{align*}
p_{n+1}(x)\equiv p\left(x,\tau+\frac{1}{k_1}\right),\quad \rho_{n+1}(x)\equiv \rho\left(x,\tau+\frac{1}{k_1}\right),\\
q_{n+1}(x)\equiv q\left(x,\tau+\frac{1}{k_1}\right),\quad \sigma_{n+1}(x)\equiv \sigma\left(x,\tau+\frac{1}{k_1}\right),
\end{align*}
in $\frac{1}{k_1}$, then  defining  a new independent temporal variable  $t$ in term of $\tau$ and $x$ such that
\begin{align*}
\partial_\tau=\partial_x-\frac{1}{12{k_1}^2}\partial_t,
\end{align*}
 we obtain in the continuous limit up to terms of order $\frac{1}{k_{1}^2}$
% \begin{subequations}\label{}
 \begin{align*}
  pp_{t}=&-pp_{xxx}+3p_{x}p_{xx}+6(p_{x}\rho-p\rho_{x})\sigma_{x},\\
  (p\rho)_{t}=&-(p\rho)_{xxx}+6(p_{x}\rho_{xx}+\rho_{x}p_{xx})
  +6(p_{x}^2-pp_{xx}+2\rho_{x}\rho)\sigma_{x}-6q_{x}(p_{x}\rho-p\rho_{x}),  \\
  \sigma_{t}=&-\sigma_{xxx}+6\left(q_{x}-\frac{p_{xx}}{p}+\frac{2p_{x}^2}{p^2}\right)\sigma_{x},\\
  q_{t}=&-q_{xxx}+6q_{x}\left(q_{x}-\frac{p_{xx}}{p}+\frac{2p_{x}^2}{p^2}\right)
  -6\sigma_{x}\left(\sigma_{xx}-\frac{\rho_{xx}}{p}+\frac{\rho p_{xx}}{p^2}
  +4\frac{p_{x}\rho_{x}}{p^2}-4\frac{p_{x}^2\rho}{p^3}\right).
 \end{align*}
%\end{subequations}
 which is nothing but the component form of potential SKdV$_{-2}$  equation (\ref{skdvcomponent}).

Now we turn to the system (\ref{nsfdiscrete}) subject to \eqref{extra}. As usual, different continuous limits may be considered in the present case. For instance, we may consider the so-called straight continuum limit.
%\subsection{The semi-continuous limits of (\ref{nsfdiscrete})}
%We present here two different results obtained by implementing different continuous limits of (\ref{nsfdiscrete}), at first when we send to infinity just one  of the discrete variables and secondly when we send to infinity a combination of both discrete variables.
%\subsubsection{Straight continuum limit}
%The system (\ref{nsfdiscrete}) may be regarded as a discrete analogue
%of the differential-difference equation (\ref{btdiscrete}).
To do so, let us assume
\begin{align*}
p_{n,m}\equiv p_{n}(x),\quad \rho_{n,m}\equiv \rho_{n}(x),
\quad q_{n,m}\equiv q_{n}(x),\quad \sigma_{n,m}\equiv \sigma_{n}(x),\quad x=\frac{m}{k_2}.
\end{align*}
For $\frac{1}{k_2}$ small, we have the following Taylor series expansions
\begin{align*}
p_{n,m+1}=p_{n}\left(x+\frac{1}{k_2}\right)&=p_{n}+\frac{1}{k_2}p_{n,x}+O\left(\frac{1}{{k_2}^2}\right),\\
\rho_{n,m+1}=\rho_{n}\left(x+\frac{1}{{k_2}}\right)&=\rho_{n}+\frac{1}{{k_2}}\rho_{n,x}+O\left(\frac{1}{{k_2}^2}\right),\\
q_{n,m+1}=q_{n}\left(x+\frac{1}{k_2}\right)&=q_{n}+\frac{1}{k_2}q_{n,x}+O\left(\frac{1}{{k_2}^2}\right),\\
\sigma_{n,m+1}=\sigma_{n}\left(x+\frac{1}{{k_2}}\right)&=\sigma_{n}+\frac{1}{{k_2}}\sigma_{n,x}+O\left(\frac{1}{{k_2}^2}\right),
\end{align*}
Substituting the above expansions into (\ref{nsfdiscrete},\ref{extra}), the leading terms yield
%\begin{subequations}\label{straight}
% \begin{align}
%(p_{n+1}p_{n})_x=&\;
%  k_{1}(p_{n+1}^2-p_{n}^2)+(\sigma_{n+1}-\sigma_{n})(p_{n}\rho_{n+1}-p_{n+1}\rho_{n}),  \\
% (p_{n+1}\rho_{n}+\rho_{n+1}p_{n})_x=&\;2k_{1}(p_{n+1}\rho_{n+1}-p_{n}\rho_{n})+(q_{n+1}-q_{n})(p_{n}
% \rho_{n+1}-p_{n+1}\rho_{n})  \nonumber\\
% &-(\sigma_{n+1}-\sigma_{n})(p_{n} p_{n+1,x}-p_{n+1}p_{n,x}+2\rho_{n}\rho_{n+1}),  \\
% (\sigma_{n+1}+\sigma_{n})_x=&\;\left(\frac{2p_{n,x}}{p_{n}}+\frac{2k_{1}p_{n}}{p_{n+1}}+q_{n+1}-q_{n}\right)(\sigma_{n+1}-\sigma_{n}),\\
% (q_{n+1}+q_{n})_x=&\;\left(\frac{2p_{n,x}}{p_{n}}+\frac{2k_1p_{n}}{p_{n+1}}+q_{n+1}-q_{n}\right)(q_{n+1}-q_{n})    \nonumber\\
% &+\left[2(\frac{\rho_{n}}{p_{n}})_x
%  +\frac{2k_{1}}{p_{n+1}}(\rho_{n}-\frac{p_{n}\rho_{n+1}}{p_{n+1}})
%  +(\sigma_{n+1}-\sigma_{n})_x\right](\sigma_{n+1}-\sigma_{n}).\nonumber\\
% \end{align}
%\end{subequations}
the system (\ref{btdiscrete}). Thus, we may claim that the  discrete system (\ref{nsfdiscrete},\ref{extra}) is a fully discrete version of the potential SKdV$_2$.

It is possible to consider other continuum limits such as skew continuum limit or full continuum limit for the system (\ref{nsfdiscrete},\ref{extra}), but we will not present the results  here since these can be done thorough  tedious but straightforward calculations.

\section*{Acknowledgments}
This paper is supported by the National Natural Science Foundation of China (grant numbers: 11271366 and 11331008) and the Fundamental Research Funds for Central Universities.

\end{document}